\documentclass[aps,twocolumn,showpacs,superscriptaddress,longbibliography]{revtex4-2}

\usepackage{graphicx}
\usepackage{bm}
\usepackage{amsmath}
\usepackage{dcolumn}%

\newcommand{\eg}{\ensuremath{e_{g}}}
\newcommand{\tg}{\ensuremath{t_{2g}}}
\newcommand{\dxy}{\ensuremath{d_{xy}}}
\newcommand{\dxz}{\ensuremath{d_{xz}}}
\newcommand{\dyz}{\ensuremath{d_{yz}}}
\newcommand{\dzz}{\ensuremath{d_{3z^2-1}}}
\newcommand{\dxxyy}{\ensuremath{d_{x^2-y^2}}}

\newcolumntype{/}{D{/}{/}{2,2}}  
\newcolumntype{.}{D{.}{.}{0}}  

\begin{document}

\title{Resonant inelastic x-ray scattering in layered trimer iridate
  Ba$_4$NbIr$_3$O$_{12}$: the density functional approach }

\author{D.A. Kukusta}

\affiliation{G. V. Kurdyumov Institute for Metal Physics of the
  N.A.S. of Ukraine, 36 Academician Vernadsky Boulevard, UA-03142
  Kyiv, Ukraine}

\author{L.V. Bekenov}

\affiliation{G. V. Kurdyumov Institute for Metal Physics of the
  N.A.S. of Ukraine, 36 Academician Vernadsky Boulevard, UA-03142
  Kyiv, Ukraine}

\author{V.N. Antonov}

\affiliation{G. V. Kurdyumov Institute for Metal Physics of the
  N.A.S. of Ukraine, 36 Academician Vernadsky Boulevard, UA-03142
  Kyiv, Ukraine}

\date{\today}

\begin{abstract}

We have investigated the electronic structure of Ba$_4$NbIr$_3$O$_{12}$ within
the density-functional theory (DFT) using the generalized gradient
approximation while considering strong Coulomb correlations (GGA+$U$) in the
framework of the fully relativistic spin-polarized Dirac linear muffin-tin
orbital band-structure method. Ba$_4$NbIr$_3$O$_{12}$ has a quasi-2D structure
composed of corner-connected Ir$_3$O$_{12}$ trimers containing three distorted
face-sharing IrO$_6$ octahedra. The Ir atoms are distributed over two
symmetrically inequivalent sites: the center of the trimer (Ir$_1$) and its
two tips (Ir$_2$).  The Ir$_1$ $-$ Ir$_2$ distance within the trimer is quite
small and equals to 2.547\AA\, at low temperature. As a result, there is clear
formation of bonding and antibonding states. The large bonding-antibonding
splitting stabilizes the {\dzz}-orbital-dominant antibonding state of 5$d$
holes and produces a wide energy gap at the Fermi level. The ground state of
Ba$_4$NbIr$_3$O$_{12}$ is a nonmagnetic singlet. Relatively moderate
spin-orbit coupling (SOC) together with molecular orbital-like states in
Ba$_4$NbIr$_3$O$_{12}$ suggest that the pure $J_{\rm{eff}}$ = 1/2 model may
not be appropriate for this oxide. We have theoretically calculated the x-ray
absorption spectroscopy (XAS) spectra at the Ir $L_{2,3}$ and Nb $L_3$ edges
as well as the photoemission spectrum of Ba$_4$NbIr$_3$O$_{12}$. We have also
presented a comprehensive investigation of the resonant inelastic x-ray
scattering (RIXS) spectra at the Ir $L_3$, O $K$, Nb $K$, $L_3$, M$_3$, $M_5$,
and $N_3$ edges. The RIXS spectrum of Ba$_4$NbIr$_3$O$_{12}$ at the Ir $L_3$
edge possesses several sharp features $\le$2 eV corresponding to transitions
within the Ir {\tg} levels. The peak located at $\sim$3.2 eV is found to be
due to $\tg \rightarrow \eg$ transitions. Some amount of $\tg \rightarrow \tg$
transitions also contribute to the intensity of this peak. The high energy
fine structure $\ge$5.3 eV is mostly determined by 5$d_{\rm{O}}$ $\rightarrow
\tg$ and O$_{2p}$ $\rightarrow$ {\eg} transitions. The spectral features
between 8 and 12 eV are due to 5$d_{\rm{O}}$ $\rightarrow \eg$ transitions. We
have also presented the investigation of the dependence of the Ir $L_3$ RIXS
spectrum on the momentum transfer vector {\bf Q} and incident photon energy
$E_i$.

\end{abstract}

\pacs{75.50.Cc, 71.20.Lp, 71.15.Rf}

\maketitle

\section{Introduction}

\label{sec:introd}

The interplay between orbital, spin, charge carriers, and lattice degrees of
freedom has been a fascinating subject for the condensed-matter physics
community for the last few decades. The 5$d$ transition-metal compounds
possess the same order of magnitude of on-site Coulomb repulsion $U$,
spin-orbit coupling (SOC), and crystal-field energy \cite{WCK+14}. So far,
most researches have largely focused on iridium oxides for novel phenomena,
such as topological insulators \cite{QZ10,Ando13,WBB14,BLD16}, Mott insulators
\cite{KJM+08,KOK+09,WSY10,MAV+11}, Weyl semimetals \cite{WiKi12,GWJ12,SHJ+15},
and quantum spin liquids (QSLs) \cite{KAV14,Bal10,SaBa17}. In a QSL, spins are
strongly correlated, but quantum fluctuations prevent them from long-range
ordering \cite{SaBa17}.

The specific physical properties of iridates strongly depend on local
geometry. The most widely discussed geometry is the one with MO$_6$ octahedra
(M is a transition-metal ion) sharing a common oxygen (a common corner) or two
common oxygens (octahedra with a common edge).  The first case takes place,
for example, in the LaMnO$_3$ perovskite or the layered system
La$_2$CuO$_4$. The situation with a common edge occurs in many layered systems
with triangular lattices such as NaCoO$_2$ and LiNiO$_2$. The features of SO
systems in both these cases were studied in detail (see, e.g.,
Ref. \cite{book:Khomski14}). However, there exists yet the third typical
geometry, which is also very often met in many real materials: the case of
octahedra with a common face (three common oxygens). This case has relatively
small attention in the literature. Actually, there are many transition-metal
compounds with the face-sharing geometry \cite{KKS+15}. Such materials
include, for example, the series of the 6H-hexagonal perovskite iridates
Ba$_3$MIr$_2$O$_9$ (M = Y$^{3+}$, Sc$^{3+}$, In$^{3+}$, Lu$^{3+}$)
\cite{SDH06,DoHi04}. These systems possess blocks of two face-sharing IrO$_6$
octahedra separated by MO$_6$ octahedra (which have common corners with
IrO$_6$). They have two equivalent Ir sites, therefore, the oxidation state of
Ir is fractional: Ba$_3$M$^{3+}$Ir$_2^{4.5+}$O$_9$.  These systems have been
extensively studied for novel quantum magnetic phases
\cite{DMK+12,DMK+13,DeMa13,DKM+14,NMB+16,KSP+16,DMO+17,NBB+18,KBN+19,NgCa21,GCN22}.

Another example with two blocks of face-sharing octahedra is the
Ba$_5$AlIr$_2$O$_{11}$ barium oxide. The crystal structure of this compound
consists of AlO$_4$ tetrahedra and IrO$_6$ octahedra. The latter share a face
and develop Ir$_2$O$_9$ dimers, which spread along the $a$ axis. Therefore,
the compound would become geometrically frustrated in the presence of
antiferromagnetic (AFM) interaction. Novel properties are expected from this
structural arrangement in addition to those driven by SOC.  Opposite to the
Ba$_3$MIr$_2$O$_9$ systems, Ba$_5$AlIr$_2$O$_{11}$ features dimer chains of
two inequivalent octahedra occupied by tetravalent Ir$^{4+}$ (5$d^5$) and
pentavalent Ir$^{5+}$ (5$d^4$) ions, respectively. Ba$_5$AlIr$_2$O$_{11}$ is a
Mott insulator that undergoes a subtle structural phase transition near $T_S$
= 210 K and a transition to magnetic order at $T_M$ = 4.5 K.

Recently, unconventional electronic and magnetic ground states have been
reported in compounds with Ir trimers, i.e., three face-sharing IrO$_6$
octahedra, which are much less explored so far
\cite{CCG+00,NgCa19,CZZ+20}. Among them, Ba$_4$Ir$_3$O$_{10}$ with a Ir 5$d^5$
nominal atomic configuration and a monoclinic $P2_1/a$ structure
\cite{CHW+21}. There are several experimental investigations of the physical
properties of Ba$_4$Ir$_3$O$_{10}$ with some controversial results and
conclusions \cite{WM91,KRD+11,SK+12,KKS+15,CZZ+20,CZH+20,CHW+21,SSF+22}.

Here we report a theoretical investigation of the electronic structure of the
Ba$_4$NbIr$_3$O$_{12}$ oxide. This perovskite belongs to the family of
trimer-based 12$L$-hexagonal quadrupole perovskite counterparts with the
general formula Ba$_4$MM$^{\prime}_3$O$_{12}$ (M = 3$d$ transition metal or
rare earth; M$^\prime$ = heavier transition metal Ir, Rh, Ru, etc.). Their
structures comprise single MO$_6$ octahedra and M$^{\prime}_3$O$_{12}$ trimers
of three face-sharing M$^{\prime}$O$_6$ octahedra. These perovskites have
received more limited attention, despite their potential to harbor novel
quantum magnetism including QSL phases \cite{BLA+24b}. Most previous studies
of Ba$_4$MM$^{\prime}_3$O$_{12}$ perovskites focused on their structure and
bulk magnetic properties \cite{SDW+09,SDW+10}.

The Ir atoms in Ba$_4$NbIr$_3$O$_{12}$ are distributed over two symmetrically
unequivalent sites: the center of the trimer (Ir$_1$) and its two tips
(Ir$_2$). The shortest Ir-Ir bond is the one between Ir$_1$ and Ir$_2$ within
the trimer. It has a length of $\sim$2.547 \AA\, \cite{CZZ+20,SSL02}, which is
shorter than in the metallic iridium ($\sim$2.71 \AA\, \cite{book:Wyck63}) as
well as in the dimer perovskites Ba$_5$AlIr$_2$O$_{11}$ ($\sim$2.73 \AA\,
\cite{MuLa89}) and Ba$_3$InIr$_2$O$_9$ ($\sim$2.65 \AA\, \cite{DMO+17}), and
the trimer oxide Ba$_4$Ir$_3$O$_{10}$ ($\sim$2.58 \AA\, \cite{CHW+21}).

Recently, Nguyen {\it et al.} investigated three trimer systems
Ba$_4$NbM$^{\prime}_3$O$_{12}$ (M$^{\prime}$ = Ir, Rh, Ru) as candidate QSL
materials \cite{NHW+18,NgCa19}. Thakur {\it et al.}  investigated the magnetic
structure of Ba$_4$Nb$_{0.8}$Ir$_{3.2}$O$_{12}$ single crystals also to reveal
the possibility of a spin-liquid state \cite{TCD+20}. Bandyopadhyay {\it at
  al.}  investigated experimentally the magnetic ground state in
Ba$_4$NbIr$_3$O$_{12}$ \cite{BLA+24b} using a diverse range of techniques
including powder x-ray diffraction (XRD), x-ray absorption spectroscopy (XAS)
and x-ray photoemission spectroscopies (XPES), electrical resistivity, dc and
ac magnetic susceptibilities, specific heat, and muon spin rotation/relaxation
($\mu$SR). The electrical resistivity together with XPS spectroscopy data as
well as {\it ab initio} electronic structure calculations suggested an
insulating electronic ground state of Ba$_4$NbIr$_3$O$_{12}$. The activated
energy gap of $\sim$0.22-0.25 eV was estimated from the temperature dependence
of the electrical resistivity \cite{BLA+24b}. A quantitative analysis of
Ir-$L_{2,3}$ XAS spectra established a moderate effective SOC strength for Ir
\cite{BLA+24b}. The Curie-Weiss temperature for Ba$_4$NbIr$_3$O$_{12}$ is
$-$13 K signaling rather small intercluster exchange
\cite{NgCa19}. Ba$_4$NbIr$_3$O$_{12}$ remains nonmagnetic down to the lowest
temperatures and was suggested to be a spin liquid
\cite{NgCa19,KKS20,BLA+24b}.

In this paper, we report a theoretical investigation from the first principles
of the RIXS spectra of Ba$_4$NbIr$_3$O$_{12}$.  The RIXS method has shown
remarkable progress as a spectroscopic technique to record the momentum and
energy dependence of inelastically scattered photons in complex
materials. RIXS rapidly became the forefront of experimental photon science
\cite{AVD+11,GHE+24}. RIXS combines spectroscopy and inelastic scattering to
probe the electronic structure of materials. This method is also bulk
sensitive, polarization dependent, as well as element and orbital specific
\cite{AVD+11}. It permits direct measurements of phonons, plasmons,
single-magnons and orbitons, as well as other many-body excitations in
strongly correlated systems, such as cuprates, nickelates, osmates,
ruthenates, and iridates, with complex low-energy physics and exotic phenomena
in the energy and momentum space.

Recently, the RIXS measurements at the Ir $L_3$ edge in Ba$_4$NbIr$_3$O$_{12}$
have been successfully performed by Magnaterra {\it et al.}  \cite{MSS+25} in
the energy range up to 8.5 eV. In addition to the elastic peak centered at
zero energy loss, the spectrum consists of several peaks below 2 eV and a peak
at 3.8 eV. The authors also presented the dependence of the RIXS spectra on
the momentum transfer vector {\bf Q} in the energy interval $\le$2.5 eV.

Our detailed study of the electronic structure and RIXS spectra of
Ba$_4$NbIr$_3$O$_{12}$ is carried out in terms of the density functional
theory. The study sheds light on the important role of band structure effects
and transition metal 5$d$ $-$ oxygen 2$p$ hybridization in the spectral
properties of 5$d$ oxides. We use the {\it ab initio} approach using the fully
relativistic spin-polarized Dirac linear muffin-tin orbital band-structure
method. The paper is organized as follows. The crystal structure of
Ba$_4$NbIr$_3$O$_{12}$ and computational details are presented in
Sec. II. Section III presents the electronic structure of
Ba$_4$NbIr$_3$O$_{12}$. In Sec. IV, the theoretical investigation of the Ir
photoemission spectrum and XAS spectra at the Ir $L_{2,3}$ and Nb $L_3$ edges
are presented. We also investigate theoretically the RIXS spectra of
Ba$_4$NbIr$_3$O$_{12}$ at the Ir $L_3$, O $K$, and Nb $K$, $L_3$, M$_3$,
$M_5$, and $N_3$ edges. The theoretical results are compared with experimental
measurements. Finally, the results are summarized in Sec. V.

\section{Computational details}
\label{sec:details}

\subsection{RIXS}  

In the direct RIXS process \cite{AVD+11} the incoming photon with energy
$\hbar \omega_{\mathbf{k}}$, momentum $\hbar \mathbf{k}$, and polarization
$\bm{\epsilon}$ excites the solid from ground state $|{\rm g}\rangle$ with
energy $E_{\rm g}$ to intermediate state $|{\rm I}\rangle$ with energy $E_{\rm
  I}$. During relaxation an outgoing photon with energy $\hbar
\omega_{\mathbf{k}'}$, momentum $\hbar \mathbf{k}'$ and polarization
$\bm{\epsilon}'$ is emitted, and the solid is in state $|f \rangle$ with
energy $E_{\rm f}$. As a result, an excitation with energy $\hbar \omega =
\hbar \omega_{\mathbf{k}} - \hbar \omega_{\mathbf{k}'}$ and momentum $\hbar
\mathbf{q}$ = $\hbar \mathbf{k} - \hbar \mathbf{k}'$ is created.  Our
implementation of the code for the calculation of the RIXS intensity uses
Dirac four-component basis functions \cite{NKA+83} in the perturbative
approach \cite{ASG97}. RIXS is a second-order process, and its intensity is
given by

\begin{eqnarray}
I(\omega, \mathbf{k}, \mathbf{k}', \bm{\epsilon}, \bm{\epsilon}')
&\propto&\sum_{\rm f}\left| \sum_{\rm I}{\langle{\rm
    f}|\hat{H}'_{\mathbf{k}'\bm{\epsilon}'}|{\rm I}\rangle \langle{\rm
    I}|\hat{H}'_{\mathbf{k}\bm{\epsilon}}|{\rm g}\rangle\over
  E_{\rm g}-E_{\rm I}} \right|^2 \nonumber \\ && \times
\delta(E_{\rm f}-E_{\rm g}-\hbar\omega),
\label{I1}
\end{eqnarray}
where the RIXS perturbation operator in the dipole approximation is given by
the lattice sum $\hat{H}'_{\mathbf{k}\bm{\epsilon}}=
\sum_\mathbf{R}\hat{\bm{\alpha}}\bm{\epsilon} \exp(-{\rm
  i}\mathbf{k}\mathbf{R})$, where $\bm{\alpha}$ are the Dirac matrices. The
sum over intermediate states $|{\rm I}\rangle$ includes the contributions
from different spin-split core states at the given absorption edge. The matrix
elements of the RIXS process in the frame of the fully relativistic Dirac LMTO
method are presented in Ref. \cite{AKB22a}.

\subsection{Crystal structure} 

\begin{figure}[tbp!]
\begin{center}
\includegraphics[width=0.90\columnwidth]{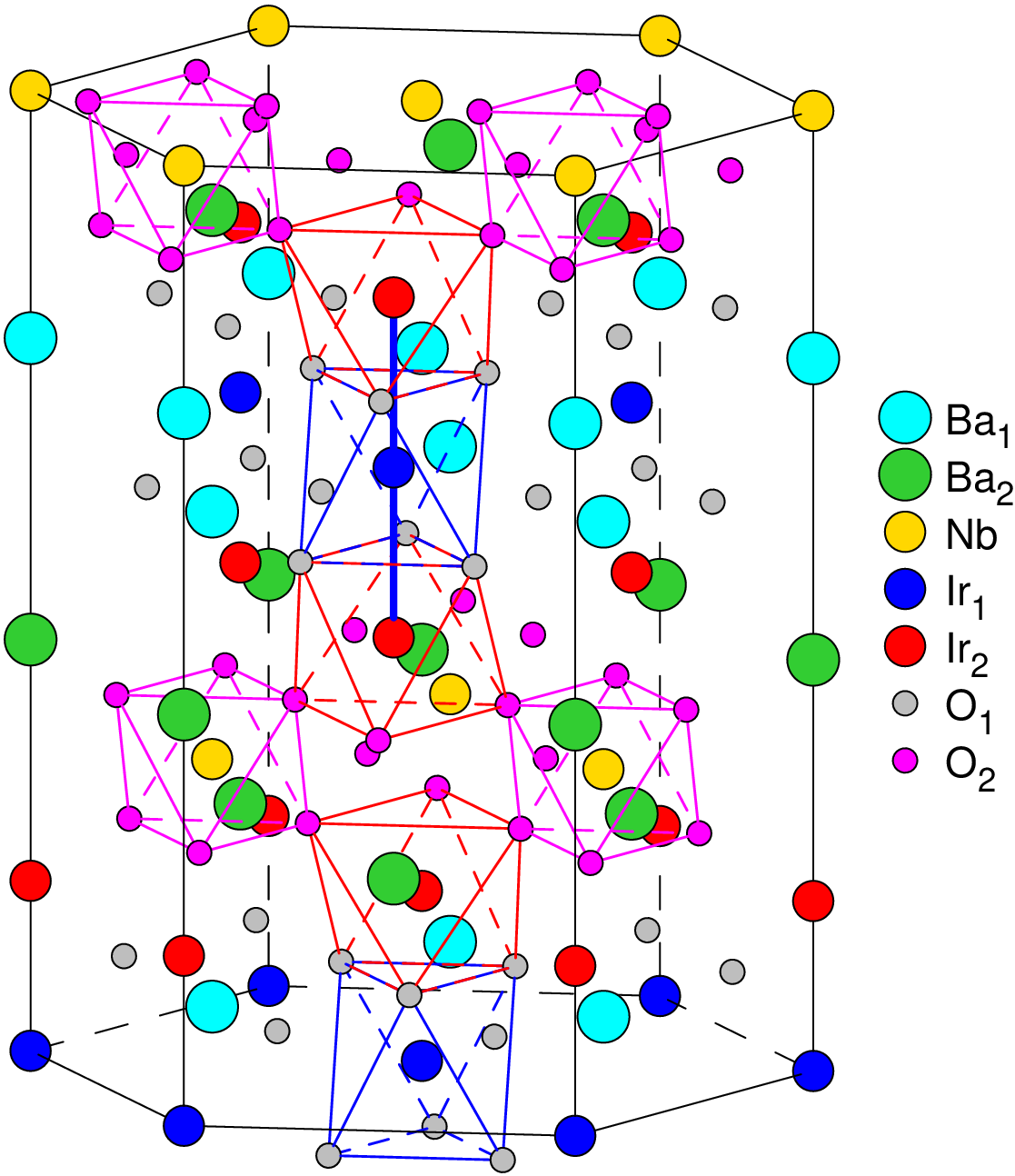}
\end{center}
\caption{\label{struc_BNIO}(Color online) The schematic representation
  of Ba$_4$NbIr$_3$O$_{12}$ in the rhombohedral structure with the
  space group $R\bar{3}m$ (No. 166). }
\end{figure}

The Ba$_4$NbIr$_3$O$_{12}$ compound was synthesized recently by L. T. Nguyen
and R. J. Cava. \cite{NgCa19}. The lattice and structural parameters are
summarized in Table \ref{struc_BNIO}. The structure consists of three IrO$_6$
octahedra connected by face sharing to form Ir$_3$O$_{12}$ trimers
(Fig. \ref{struc_BNIO}). The trimers are arranged in a triangular planar
lattice. While Ba$_4$LnIr$_3$O$_{12}$ (Ln = lanthanides) has the monoclinic
$C2/m$ space group, Ba$_4$NbIr$_3$O$_{12}$ adopts a higher-symmetry
rhombohedral $R\bar{3}m$ (No. 166) crystal structure \cite{NgCa19}.  The
individual Ir$_3$O$_{12}$ trimers share corners with nonmagnetic NbO$_6$
octahedra, and not with other trimers.  The NbO$_6$ octahedra and
Ir$_3$O$_{12}$ trimers alternate along the $c$ axis to generate a 12-layer
(i.e., three layers of four octahedra) hexagonal perovskite structure.  The
unit cell of Ba$_4$NbIr$_3$O$_{12}$ contains two inequivalent Ba sites, two
inequivalent O sites, and two inequivalent Ir sites \cite{NgCa19}. The two Ir
sites, Ir$_1$ and Ir$_2$, are located in the middle and outer positions of the
trimer, respectively.  In each trimer the two outside octahedra around Ir$_2$
ions are strongly distorted. The middle octahedron around the Ir$_1$ ion is
less distorted.  The Ir$_1$-O$_1$ distances are equal to 1.96727 \AA, while
the Ir$_2$-O$_1$ and Ir$_2$-O$_2$ distances are equal to 1.93698 and 2.14923
\AA, respectively. The length of the shortest Ir-Ir bond, which is the one
between Ir$_1$ and Ir$_2$ within the trimers, equals to 2.54694 \AA.

\begin{table}[tbp!]
  \caption {The Wyckoff positions (WP) of Ba$_4$NbIr$_3$O$_{12}$ for
    the rhombohedral $R\bar{3}m$ (No. 166) crystal structure (lattice
    constants $a$ = 5.7827 \AA and $c$ = 28.7725 \AA\,
    \cite{NgCa19}). }
\label{struc_tab_BNIO}
\begin{center}
\begin{tabular}{|c|c|c|c|c|c|}
\hline
Structure      & WP & Atom     & $x$      & $y$    & $z$     \\
\hline
                   & $6c$   & Ba$_1$    &  0       & 0      & 0.12890 \\
                   & $6c$   & Ba$_2$    &  0       & 0      & 0.28585 \\
                   & $3a$   & Nb        &  0       & 0      & 0      \\
$R\bar{3}m$        & $3b$   & Ir$_1$    &  0       & 0      & 0.5      \\
Ref. \cite{NgCa19} & $6c$   & Ir$_2$    &  0       & 0      & 0.41148  \\
                   & $18h$  & O$_1$     &  0.48170  & 0.518309& 0.12160   \\
                   & $18h$  & O$_2$     &  0.51460  & 0.48540 & 0.29560  \\
\hline
\end{tabular}
\end{center}
\end{table}

\subsection{Calculation details}

The details of the computational method are described in our previous papers
\cite{AJY+06,AHY+07b,AYJ10,AKB22a} and here we only mention several
aspects. The band structure calculations were performed using the fully
relativistic LMTO method \cite{And75,book:AHY04}. This implementation of the
LMTO method uses four-component basis functions constructed by solving the
Dirac equation inside an atomic sphere \cite{NKA+83}. The exchange-correlation
functional of a GGA-type was used in the version of Perdew, Burke and
Ernzerhof \cite{PBE96}. The Brillouin zone integration was performed using the
improved tetrahedron method \cite{BJA94}. The basis consisted of Nb, Ir, and
Ba $s$, $p$, $d$, and $f$; and O $s$, $p$, and $d$ LMTO's.

In the RIXS process, an electron is promoted from a core level to an
intermediate state, leaving a core hole. As a result, the electronic structure
of this state differs from that of the ground state. To reproduce the
experimental spectrum, the self-consistent calculations should be carried out
including a core hole. Usually, the core-hole effect has no impact on the
shape of XAS at the $L_{2,3}$ edges of 5$d$ systems \cite{book:AHY04}.
However, the core hole has a strong effect on the RIXS spectra in transition
metal compounds \cite{AKB22a,AKB22b}; therefore, we consider it in our
calculations.

In our calculations, we rely on the experimentally
measured atomic positions and lattice constants, because they are well
established for this material and still more accurate than those
obtained from DFT.

\section{Electronic structure}
\label{sec:bands}

\begin{figure}[tbp!]
\begin{center}
\includegraphics[width=0.8\columnwidth]{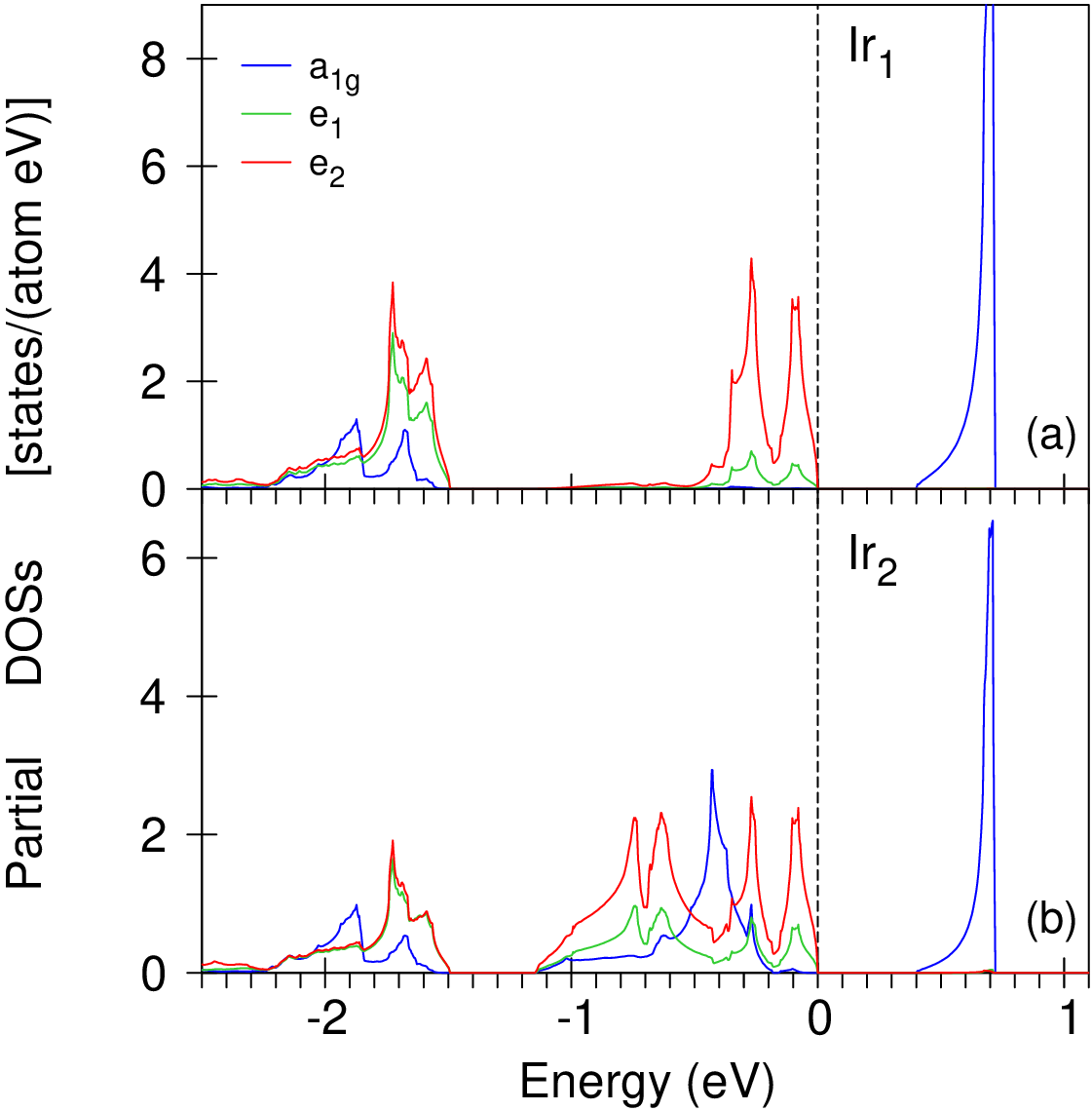}
\end{center}
\caption{\label{Orbitals_BNIO}(Color online) The orbital-resolved Ir
  {\tg} partial density of states (DOS) [in states/(atom eV)] for
  Ba$_4$NbIr$_3$O$_{12}$ calculated in the GGA approach at the
  Ir$_1$ (a) and Ir$_2$ (b) sites. }
\end{figure}

We performed GGA, GGA+SO, and GGA+SO+$U$ calculations of the electronic
structure of Ba$_4$NbIr$_3$O$_{12}$. The crystal field at the Ir site ($D3d$
and $C3v$ point symmetry for Ir$_1$ and Ir$_2$ sites, respectively) causes the
splitting of 5$d$ orbitals into one singlet $a_{1g}$ ({\dzz}), and two
doublets $e_1$ ({\dxz} and {\dyz}) and $e_2$ ({\dxxyy} and {\dxy}). The
$a_{1g}$ orbitals are oriented head-on in the face sharing geometry. In Fig.
\ref{Orbitals_BNIO} the plots of orbital-resolved partial density of states
(DOS) for Ba$_4$NbIr$_3$O$_{12}$ at the Ir$_1$ and Ir$_2$ sites calculated in
the GGA approach are presented. Both the Ir$_1$ and Ir$_2$ ions possess strong
empty DOS peaks at $\sim$0.7. There is an energy gap between the antibonding
$a_{1g}$ and $e_2$ states. The partial DOS is quite different for the two
nonequivalent Ir sites.

Every Ir$_1$ ion has two neighboring Ir ions in its close vicinity at
$\sim$2.547 \AA\, \cite{NgCa19}. However, Ir$_2$ ions have only one
neighboring Ir ion at the same distance in the trimer structure. As a result,
the Ir$_1$ ions have a tendency of dimerization and of forming quasi-molecular
orbital (QMO) states. The $a_{1g}$ ({\dzz}) orbital provides the dominant
character of the lowest ($-$1.9 eV) occupied subband and the highest (+0.7 eV)
empty subband. These two subbands with predominant {\dzz} character can be
assigned to the bonding and antibonding states of Ir$_1$ dimer molecules with
a splitting energy of $\sim$2.6 eV. The large bonding-antibonding splitting
stabilizes the {\dzz}-orbital-dominant antibonding state and produces a wide
energy gap at the Fermi level. The Ir$_2$ site also possesses the bonding and
antibonding $a_{1g}$ ({\dzz}) states at the same positions, but in addition
there are nonbonding $a_{1g}$ states at $-$0.4 eV [see
  Fig. \ref{Orbitals_BNIO}(b)], which are absent for the Ir$_1$ site. It is
important to note that the energy gap in Ba$_4$NbIr$_3$O$_{12}$ is already
opened in the GGA or GGA+SO approaches without any Hubbard $U$. Besides, our
self-consistent calculations always produce nonmagnetic (NM) solutions for
Ba$_4$NbIr$_3$O$_{12}$. Even if we start from a magnetic spin-polarized
situation, our calculations (in the GGA, GGA+SO and GGA+SO+$U$ approaches)
converge to a nonmagnetic solution.  It is in agreement with experimental
observations, which claim that Ba$_4$NbIr$_3$O$_{12}$ remains nonmagnetic down
to the lowest temperatures \cite{KKS20}. So, the ground state of
Ba$_4$NbIr$_3$O$_{12}$ is just a nonmagnetic singlet and this oxide can be
considered as a nonmagnetic noncorrelated band insulator. From this point of
view, the absence of magnetic ordering in Ba$_4$NbIr$_3$O$_{12}$ could be not
due to the spin-liquid state anticipated in Ref. \cite{NgCa19,BLA+24b} but
just due to the formation of such nonmagnetic QMO state.

\begin{figure}[tbp!]
\begin{center}
\includegraphics[width=0.98\columnwidth]{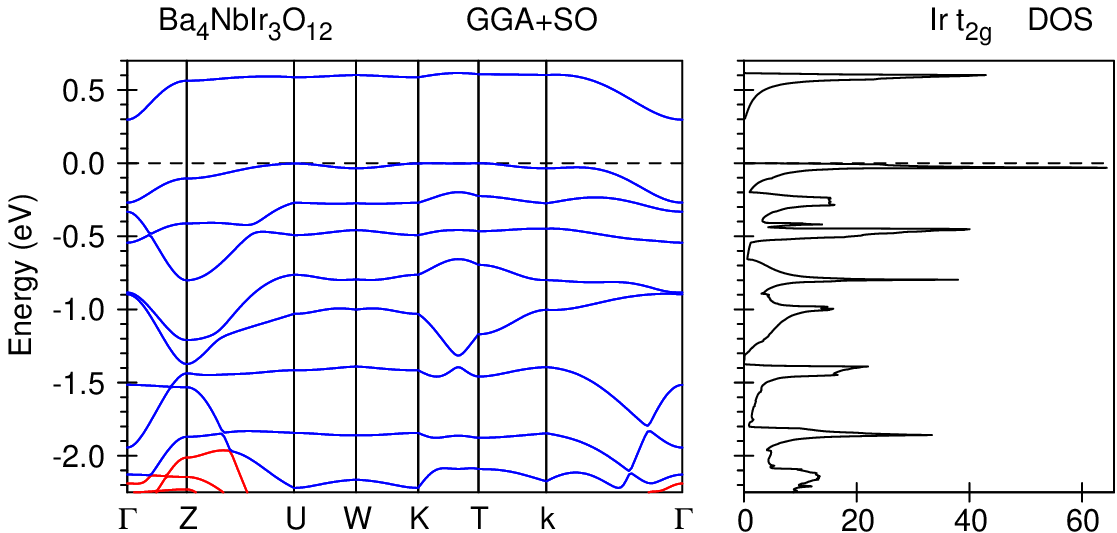}
\end{center}
\caption{\label{BND_t2g_BNIO}(Color online) The energy band structure
  and total density of states (DOS) [in states/(cell eV)] for {\tg}
  states in Ba$_4$NbIr$_3$O$_{12}$ calculated in the GGA+SO
  approach. }
\end{figure}

In Ba$_4$NbIr$_3$O$_{12}$, the inter-trimer-couplings are very small, so the
electrons are localized within the trimers, resulting in dielectric
behavior. This is supported by the band structure shown in
Fig. \ref{BND_t2g_BNIO}, where the Ir {\tg} bands calculated in the GGA+SO
approach are very narrow and avoid crossing one another. In other words, they
look just like the energy levels of molecular orbitals.

\begin{figure}[tbp!]
\begin{center}
\includegraphics[width=0.99\columnwidth]{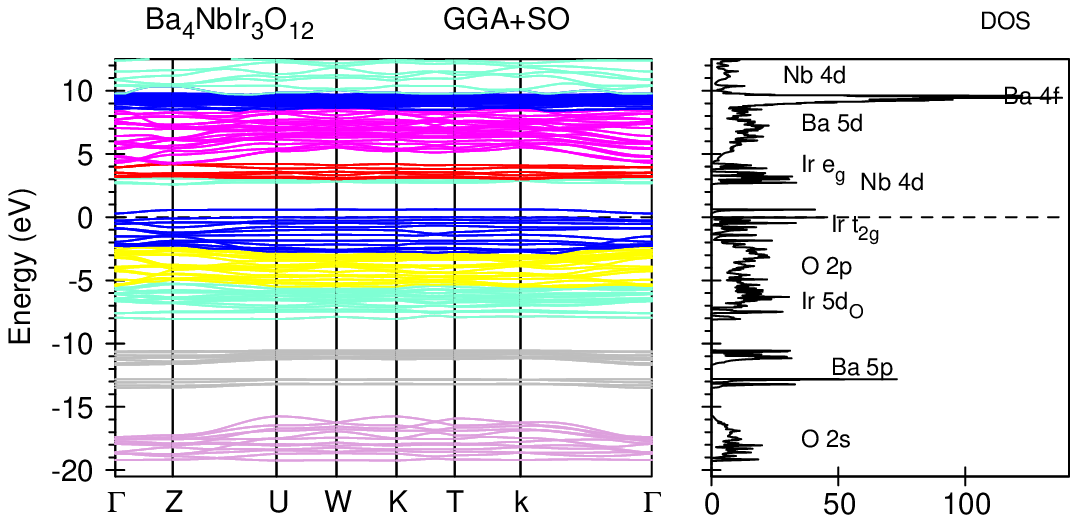}
\end{center}
\caption{\label{BND_BNIO}(Color online) The energy band structure and
  total density of states (DOS) [in states/(cell eV)] for
  Ba$_4$NbIr$_3$O$_{12}$ calculated in the GGA+SO approach. }
\end{figure}

\begin{figure}[tbp!]
\begin{center}
\includegraphics[width=0.99\columnwidth]{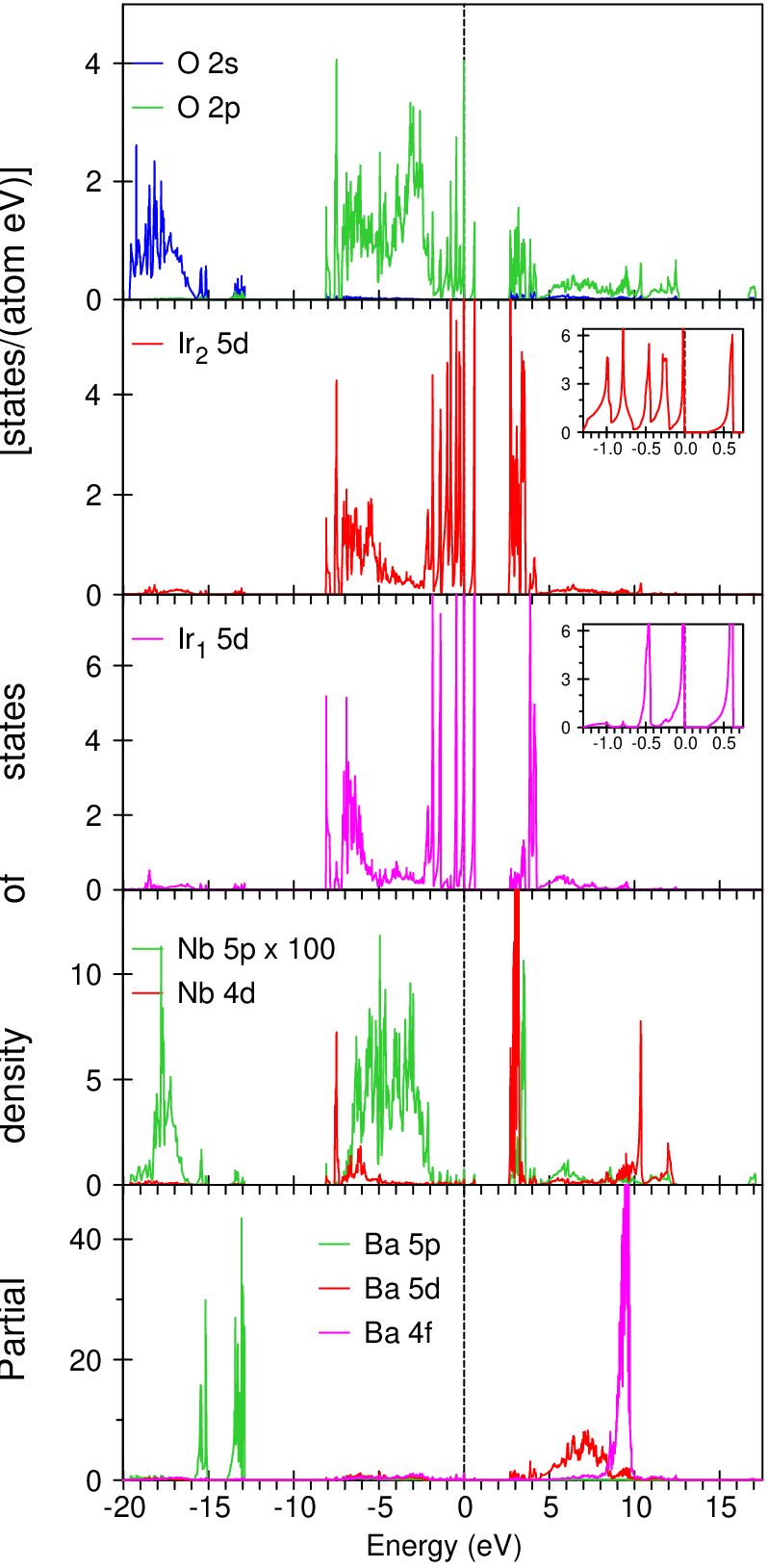}
\end{center}
\caption{\label{PDOS_BNIO}(Color online) The partial density of
  states [in states/(atom eV)] for Ba$_4$NbIr$_3$O$_{12}$
  calculated in the GGA+SO approach. }
\end{figure}

Figures \ref{BND_BNIO} and \ref{PDOS_BNIO} present the energy band structure
and partial DOSs, respectively, in Ba$_4$NbIr$_3$O$_{12}$ in the GGA+SO
approach. The occupied {\tg} states are situated in the energy interval from
$-$0.85 eV to $E_F$. There is a significant amount of Ir 5$d$ DOS located at
the bottom of the oxygen 2$p$ states from $-$7.1 to $-$5.2 eV below the Fermi
energy. These so called Ir 5$d_{\rm{O}}$ states are provided by the tails of
oxygen 2$p$ states inside the Ir atomic spheres and play an essential role in
the photo-emission spectrum as well as in the RIXS spectrum at the Ir $L_3$
edge (see Section IV). The occupied Nb 4$d$ states possess a two-peak
structure with a narrow peak at $-$7.5 eV and a wide peak from $-6$ to $-$5.7
eV. The empty Nb 4$d$ states have a strong peak from 2.6 to 3.3 eV and two
smaller peaks at 10.4 and 12 eV. It is interesting to note that the empty Nb
4$d$ states are hybridized mostly with Ir$_2$ 5$d$ states but not with Ir$_1$
5$d$ states because the intersite Nb$-$Ir$_2$ distance is much smaller than
the corresponding Nb$-$Ir$_1$ distance (4.0252 \AA\, and 5.8432 \AA,
respectively). The Nb 5$p$ occupied states are well hybridized with oxygen
2$s$ and 2$p$ states from $-$18.2 to $-$16.1 eV and from $-$6.4 to $-$2 eV,
respectively. The Ba 5$d$ states occupy the energy region from 4.8 to 8.5 eV
above the energy Fermi. A narrow and intensive DOS peak of Ba 4$f$ states is
located just above the Ba 5$d$ states from 8.7 to 9.9 eV. The Ba 5$p$
quasi-core level has two DOS peaks split by SOC corresponding to 5$p_{1/2}$
and 5$p_{3/2}$ states. There is small hybridization of Ba 5$p$ states with Nb
5$p$ and oxygen 2$s$ states. The oxygen 2$s$ states are situated far below the
Fermi level from $-$19.6 to $-$15.2 eV. The occupied O 2$p$ states are
localized from $-$8.1 eV to $E_F$. They are strongly hybridized with Ir 5$d$
states between $-$2.2 eV and $E_F$. The empty oxygen 2$p$ states are strongly
hybridized with Ir {\tg} states just above the Fermi level and with Ir {\eg}
states. They are also hybridized with Ba 5$d$ and 4$f$ states.

\section{PES, XAS, AND RIXS SPECTRA}

\subsection{I\lowercase{r} photoemission spectra}
\label{sec:pes}

\begin{figure}[tbp!]
\begin{center}
\includegraphics[width=0.9\columnwidth]{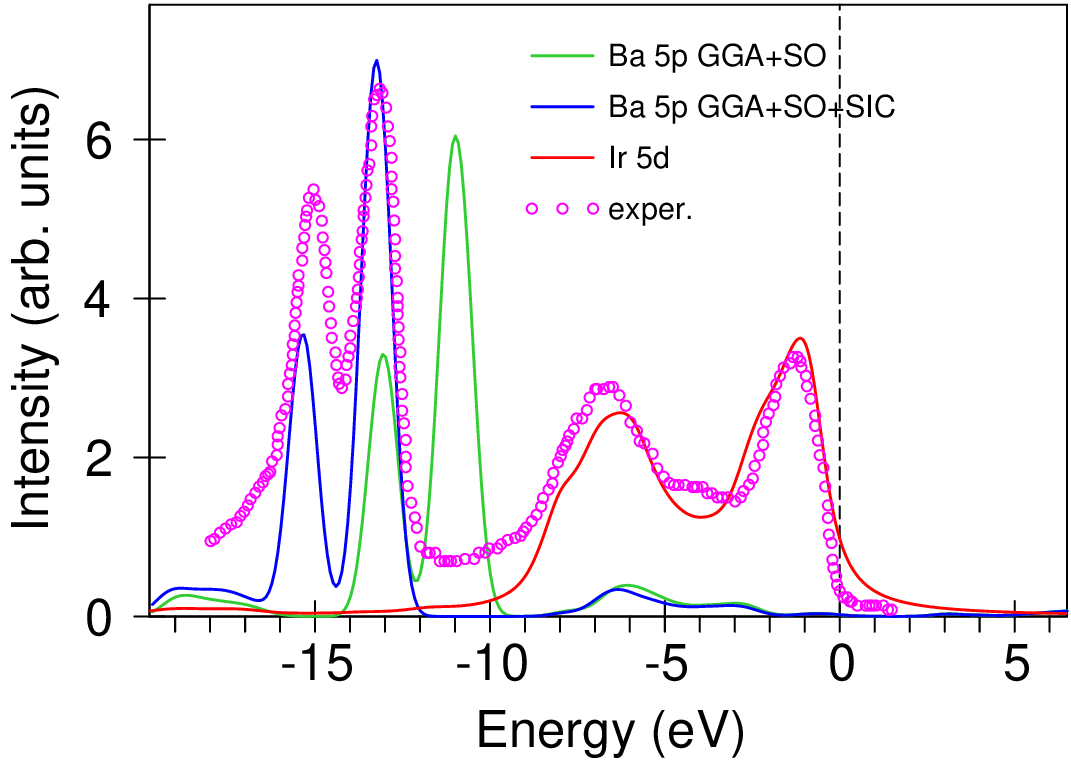}
\end{center}
\caption{\label{PES_BNIO}(Color online) The experimentally measured by
  Bandyopadhyay {\it et al.} \cite{BLA+24b} valence band photoemission
  spectrum for Ba$_4$NbIr$_3$O$_{12}$ (open magenta circles) compared with the
  Ir 5$d$ (the red curve) partial DOS calculated in the GGA+SO approach and Ba
  5$p$ quasi-core partial DOS calculated in the GGA+SO (the green curve) and
  GGA+SO+SIC (the blue curve) approaches. The partial DOSs have been
  multiplied by their photoionization cross sections. }
\end{figure}

Figure \ref{PES_BNIO} shows the experimentally measured by Bandyopadhyay {\it
  et al.} \cite{BLA+24b} valence band photoemission spectrum (PES) for
Ba$_4$NbIr$_3$O$_{12}$ (open magenta circles) compared with the Ir 5$d$ (the
red curve) and Ba 5$p$ quasi-core PDOS (the green curve) calculated in the
GGA+SO approach. We obtain the PES spectrum from the PDOS by multiplying the
PDOS by its photoionization cross section as well as the Fermi function, and
then broadening the result of multiplication. The peak situated between $-$4
eV and $E_F$ reflects the Ir {\tg} states for both Ir$_1$ and Ir$_2$ ions.
The next peak from $-$9 to $-$4 eV represents the energy distribution of the
Ir 5$d_{\rm{O}}$ states provided by the tails of oxygen 2$p$ states inside the
Ir atomic spheres and situated at the bottom of the oxygen 2$p$ states from
$-$7.1 to $-$5.2 eV below the Fermi energy (see Fig. \ref{PDOS_BNIO}). Our
GGA+SO calculations provide good agreement in the energy position and the
relative intensity of these two high energy peaks. The two deep peaks in PES
between $-$16.5 and $-$12 eV reflect the Ba 5$p$ quasi-core level splitted by
SOC into the 5$p_{1/2}$ and 5$p_{3/2}$ contributions. The GGA+SO approach was
not able to reproduce the correct energy position of the Ba 5$p$ quasi-core
levels shifting them toward the Fermi level (the green curve in
Fig. \ref{PES_BNIO}). A similar description was observed also in the
theoretical calculations by Bandyopadhyay {\it et al.}  \cite{BLA+24b} (see
Fig. 4c in their publication). To reproduce the experimental energy position
of the Ba 5$p$ states we had to take into account strong correlations of the
Ba 5$p$ states. We used a self-interaction-like correction (SIC) procedure
\cite{PZ81,LMW+10}, where the valence bands are shifted downwards by adding to
the Hamiltonian a SIC-like orbital-dependent potential $V_l$. We used the
value of $V_l$ = 2.5 eV in our band structure calculations applied to the Ba
5$p$ orbitals. The GGA+SO+SIC approach provides excellent agreement with the
experiment (see the blue curve in Fig. \ref{PES_BNIO}). A similar method was
used in the calculations of the energy band structure, optical and RIXS
spectra in Ta$_2$NiSe$_5$. To obtain a correct energy gap and experimental
optical and RIXS spectra in this chalcogenide researchers applied $V_l$ of the
same value to the Se 4$p$ states \cite{KTK+13,LYP+17,KBY+25}.

\subsection{I\lowercase{r} and N\lowercase{b} $L_{2,3}$ XAS spectra}
\label{sec:xmcd}

\begin{figure}[tbp!]
\begin{center}
\includegraphics[width=0.99\columnwidth]{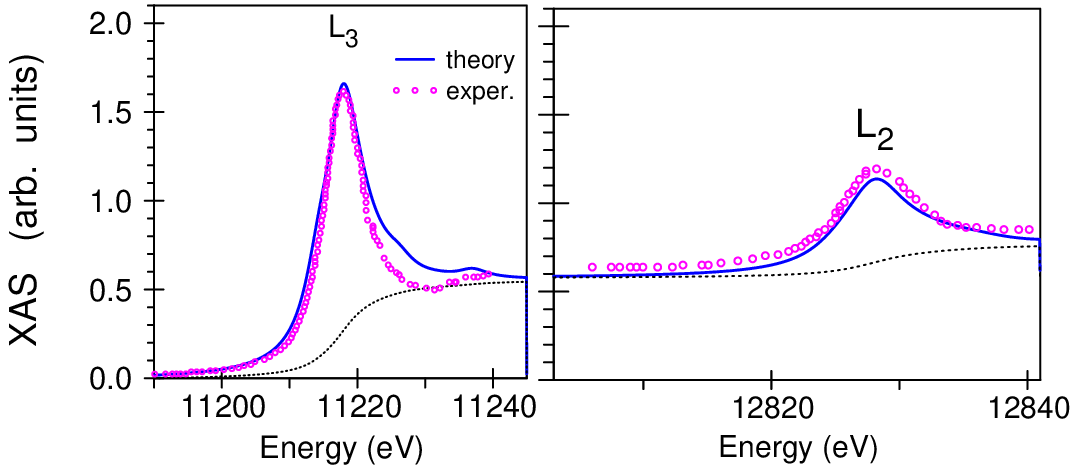}
\end{center}
\caption{\label{XAS_Ir_BNIO}(Color online) The experimentally measured
  by Bandyopadhyay {\it et al.} \cite{BLA+24b} x-ray absorption
  spectra at the Ir $L_{2,3}$ edges in Ba$_4$NbIr$_3$O$_{12}$ (open
  magenta circles) in comparison with the ones theoretically calculated in
  the GGA+SO approach. The dotted black curves
  show the background scattering intensity. }
\end{figure}

Figure \ref{XAS_Ir_BNIO} presents the experimentally measured \cite{BLA+24b}
x-ray absorption spectra at the Ir $L_{2,3}$ edges in Ba$_4$NbIr$_3$O$_{12}$
in comparison with the ones theoretically calculated in the GGA+SO
approach. There is good agreement between the calculated and experimental
spectra. The isotropic XAS spectra are mostly dominated by empty $e_g$ states
with a smaller contribution from empty $t_{2g}$ orbitals at lower energy.

The XAS spectra can provide a direct probe of SOC through a branching ratio BR
= $I_{L_3}/I_{L_2}$. In the limit of negligible SOC effects, the statistical
branching ratio BR = 2, and the $L_3$ white line is twice the size of the
$L_2$ feature \cite{LaTh88}. The theoretically calculated BR in
Ba$_4$NbIr$_3$O$_{12}$ is equal to 2.72 for the Ir$_1$ site and 2.95 for the
Ir$_2$ site. The branching ratio for the experimentally measured XAS spectra
BR $\approx$ 3.11 \cite{BLA+24b}. It is smaller than for strongly spin-orbit
coupled iridates, such as Sr$_2$IrO$_4$, for which theory gives BR = 3.56
\cite{AKB24a} and experiment gives BR = 4.1 \cite{HFZ+12}. The combined
effects of noncubic crystal distortion, Ir-O covalency, strong hybridization
resulting from direct orbital overlap of neighboring Ir-5$d$ states, intersite
Ir-Ir direct hopping, and a large bandwidth suppress the effective SOC
strength on Ba$_4$NbIr$_3$O$_{12}$. Relatively moderate SOC together with
molecular orbital-like states in Ba$_4$NbIr$_3$O$_{12}$ suggest that the pure
$J_{\rm{eff}}$ = 1/2 model may not be appropriate for this oxide.  It is
better use a description based on molecular orbital-like states. There are
several other examples in the literature of covalency-driven collapse of the
$J_{\rm{eff}}$ = 1/2 model in dimer/trimer-based iridium oxide systems
consisting of face-sharing Ir-O octahedra \cite{YKK+18,NMS+19,WWK+19,RSM+22}
as well as in the high pressure $\beta$-Li$_2$IrO$_3$ iridate
\cite{AUU18,TKG+19,AKU+21}.

\begin{figure}[tbp!]
\begin{center}
\includegraphics[width=0.9\columnwidth]{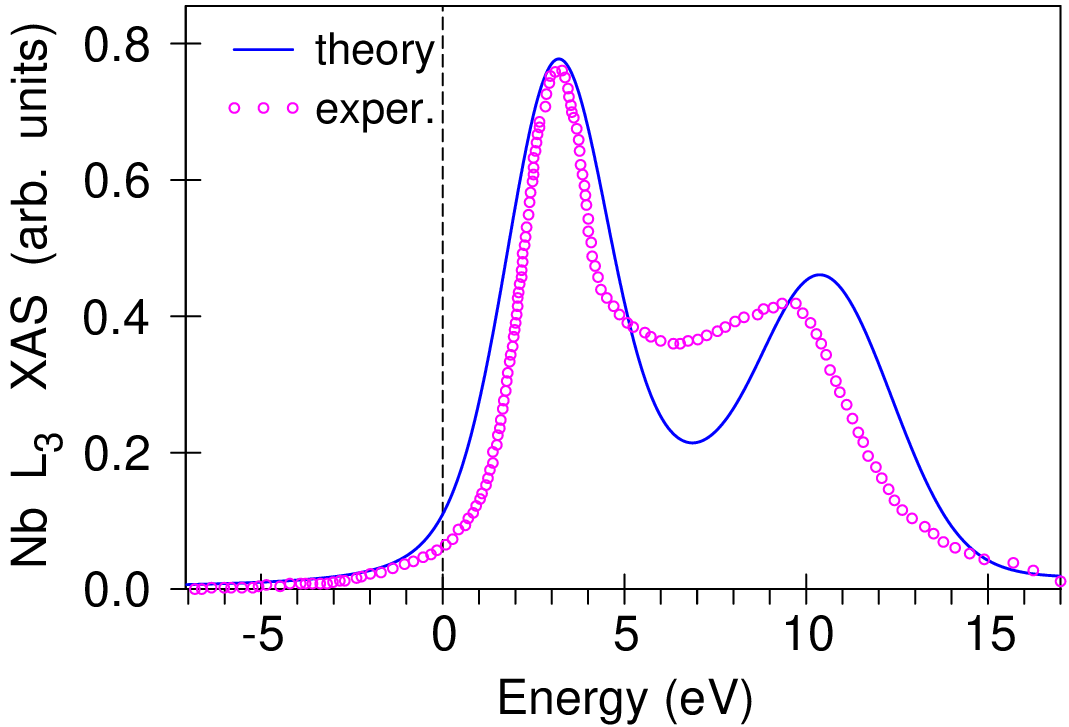}
\end{center}
\caption{\label{XAS_Nb_BNIO}(Color online) The experimentally measured
  by Bandyopadhyay {\it et al.} \cite{BLA+24b} x-ray absorption
  spectrum at the Nb $L_3$ edge in Ba$_4$NbIr$_3$O$_{12}$ (open
  magenta circles) in comparison with the one theoretically calculated in
  the GGA+SO approach. }
\end{figure}

Figure \ref{XAS_Nb_BNIO} presents the experimentally measured \cite{BLA+24b}
x-ray absorption spectrum at the Nb $L_3$ edge in Ba$_4$NbIr$_3$O$_{12}$ (open
magenta circles) in comparison with the one theoretically calculated in the
GGA+SO approach. The spectrum is due to 2$p_{3/2}$ $\rightarrow$
4$d_{3,2.5/2}$ transitions and possesses two peaks. The low energy peak
reflects the transitions from the core level into the strong narrow empty DOS
peak situated at 2.6$-$3.3 eV. The high energy peak is due to the transitions
into the two smaller DOS peaks at 10.4 and 12 eV (see
Fig. \ref{PDOS_BNIO}). The theory reproduces well the relative intensities of
these two peaks but shifts the second peak towards higher energy.

\subsection{I\lowercase{r}, O, and N\lowercase{b} RIXS spectra}
\label{sec:rixs}

\paragraph{\bf{Ir $L_3$ RIXS spectrum}}

\begin{figure}[tbp!]
\begin{center}
\includegraphics[width=0.9\columnwidth]{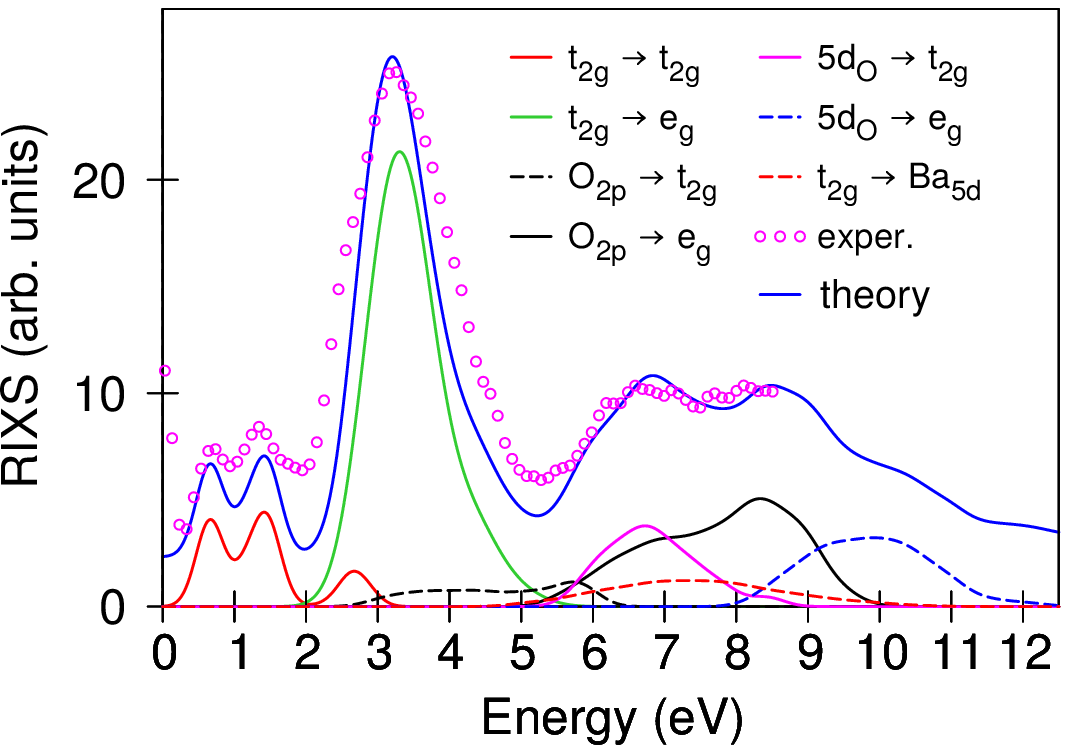}
\end{center}
\caption{\label{rixs_Ir_L3_BNIO}(Color online) The experimental
  resonant inelastic x-ray scattering (RIXS) spectrum (open magenta
  circles) measured by Magnaterra {\it et al.} \cite{MSS+25} at the Ir
  $L_3$ edge at 300 K for incident energy of 11215 eV and the momentum
  transfer vector {\bf Q} = (0.7, 0, 36.7) in reciprocal lattice units
  (r.l.u.) for the rhombohedral $R\bar{3}m$ crystal structure of
  Ba$_4$NbIr$_3$O$_{12}$ compared with the one theoretically calculated
  in the GGA+SO approach for the same incident energy and momentum transfer
  vector {\bf Q}. Calculated partial contributions from different interband
  transitions are also presented. }
\end{figure}

The experimental RIXS spectrum at the Ir $L_3$ edge was measured by Magnaterra
{\it et al.} \cite{MSS+25} in the energy range up to 8.5 eV. In addition to
the elastic peak centered at zero energy loss, the spectrum consists of two
peaks below 2 eV, a strong peak at 3.2 eV, and a wide fine structure $\ge$5.3
eV. Figure \ref{rixs_Ir_L3_BNIO} presents the theoretically calculated (the
full blue curve) and experimentally measured (open magenta circles) RIXS
spectra at the Ir $L_3$ edge for Ba$_4$NbIr$_3$O$_{12}$ \cite{MSS+25}. We also
present calculated partial contributions from different transitions between
the energy bands. We found that the fine structure situated below 2 eV
corresponds to intra-{\tg} excitation (the red curve in
Fig. \ref{rixs_Ir_L3_BNIO}). The peak located at $\sim$3.2 eV was found to be
due to $\tg \rightarrow \eg$ transitions (the green curve). Some amount of
$\tg \rightarrow \tg$ transitions also contribute to the intensity of this
peak. The two peak high energy fine structure $\ge$5.3 eV is mostly determined
by 5$d_{\rm{O}}$ $\rightarrow \tg$ (the magenta curve) and O$_{2p}$
$\rightarrow$ {\eg} (the black curve) transitions. The transitions O$_{2p}$
$\rightarrow$ {\tg} (the dashed black curve) and $\tg$ $\rightarrow$
Ba${_{5d}}$ (the dashed red curve) are relatively weak. The spectral features
between 8 and 12 eV are due to 5$d_{\rm{O}}$ $\rightarrow \eg$ transitions
(the dashed blue curve).

\begin{figure}[tbp!]
\begin{center}
\includegraphics[width=0.9\columnwidth]{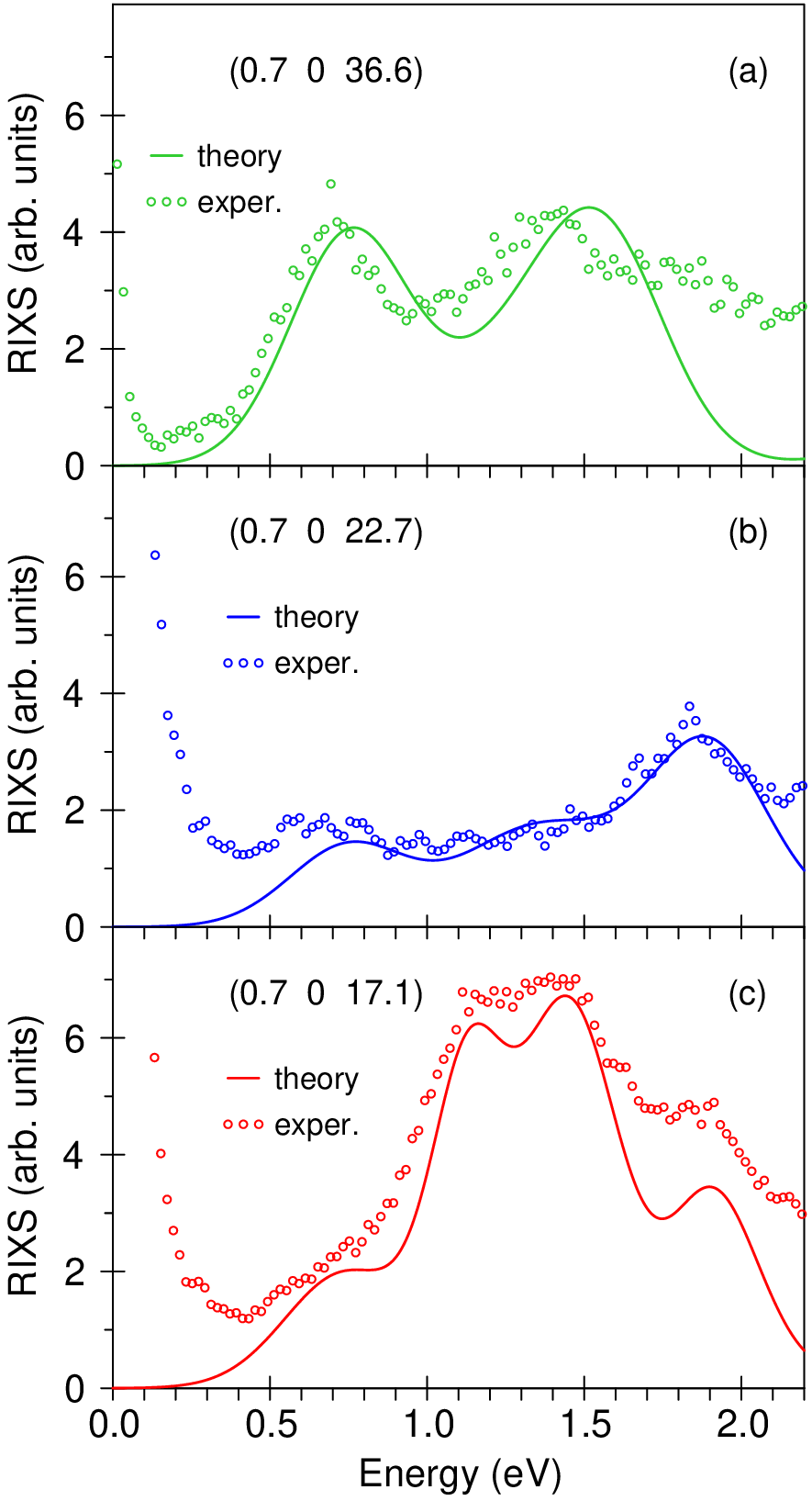}
\end{center}
\caption{\label{rixs_Qz_BNIO}(Color online)  The experimental resonance
  inelastic x-ray scattering (RIXS) spectra (open magenta circles)
  measured by Magnaterra {\it et al.} \cite{MSS+25} at the Ir $L_3$
  edge at 300 K for the momentum transfer vector {\bf Q} = (0.7, 0,
  Q$_z$) for selected values of Q$_z$ in reciprocal lattice units
  (r.l.u.) compared with the ones theoretically calculated in the GGA+SO
  approach. }
\end{figure}

\begin{figure}[tbp!]
\begin{center}
\includegraphics[width=0.9\columnwidth]{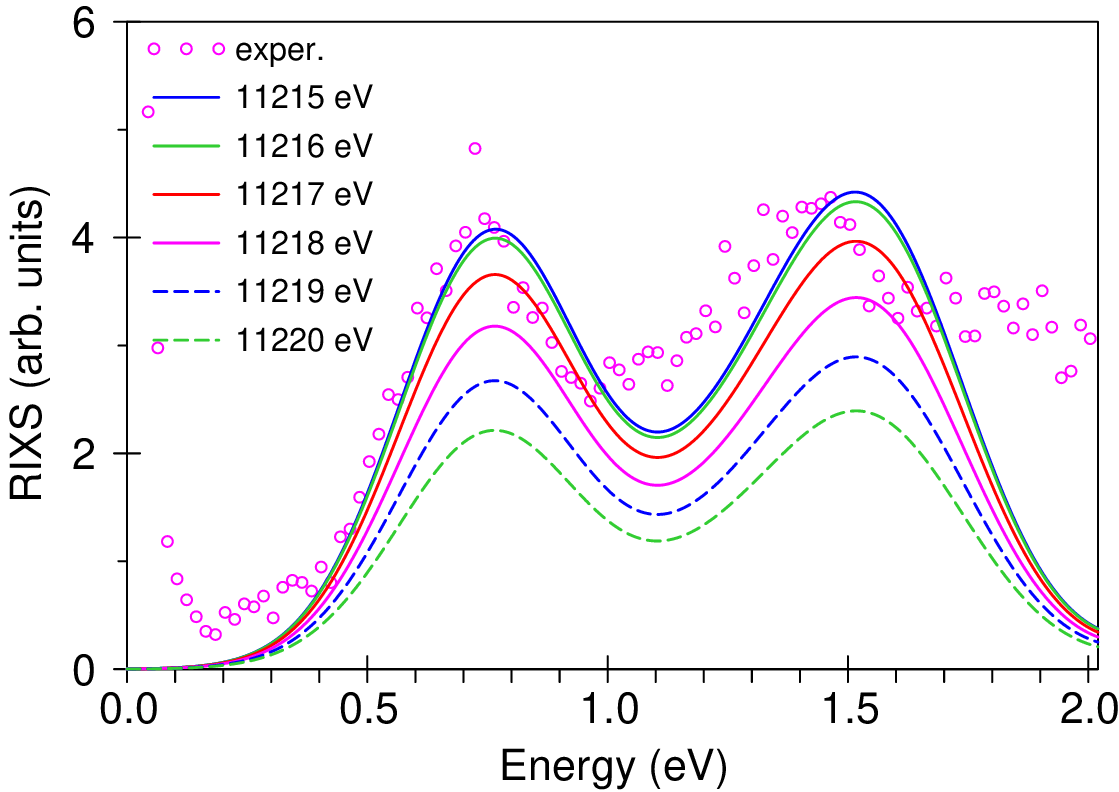}
\end{center}
\caption{\label{rixs_Ei_BNIO}(Color online) The theoretically calculated
  resonant inelastic x-ray scattering (RIXS) spectra at the Ir $L_3$ edge in
  Ba$_4$NbIr$_3$O$_{12}$ for the momentum transfer vector {\bf Q} = (0.7, 0,
  36.6) in reciprocal lattice units as a function of incident photon energy
  $E_i$. }
\end{figure}

It is widely believed that $d-d$ excitations show only small momentum transfer
vector {\bf Q} dependence in 5$d$ transition metal compounds
\cite{LKH+12,KTD+20}. Figure \ref{rixs_Qz_BNIO} shows the experimental
resonant inelastic x-ray scattering (RIXS) spectra (open circles) measured by
Magnaterra {\it et al.} \cite{MSS+25} at the Ir $L_3$ edge at 300 K for the
transferred momentum {\bf Q} = (0.7, 0, Q$_z$) parallel to the trimer axis for
selected values of Q$_z$ in reciprocal lattice units (r.l.u.)  compared with
the ones theoretically calculated in the GGA+SO approach. There is relatively
large {\bf Q} dependence of the RIXS spectrum at the Ir $L_3$ edge with
pronounced periodic modulation of the RIXS intensity as a function of
Q$_z$. This provides additional evidence for the quasi-molecular character of
the electronic structure of Ba$_4$NbIr$_3$O$_{12}$
\cite{RSM+19,RSM+22,MAP+24,MSS+25}. The theory describes quite well the
dependence of the RIXS spectrum on Q$_z$. The high energy fine structures
$>$2.5 eV are changed insignificantly with Q$_z$.

\begin{figure}[tbp!]
\begin{center}
\includegraphics[width=0.9\columnwidth]{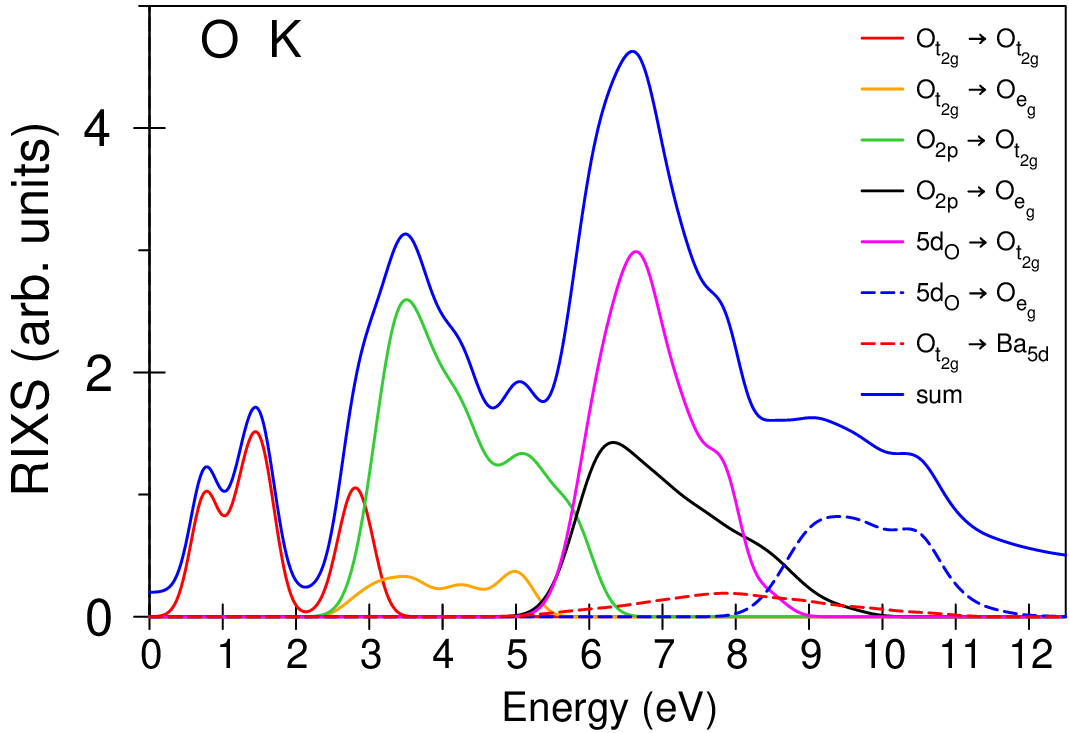}
\end{center}
\caption{\label{rixs_O_K_BNIO}(Color online) The theoretically
  calculated resonant inelastic x-ray scattering (RIXS) spectrum at
  the O $K$ edge in Ba$_4$NbIr$_3$O$_{12}$ and partial contributions
  from different interband transitions. }
\end{figure}

Figure \ref{rixs_Ei_BNIO} shows the Ir $L_3$ RIXS spectrum for the momentum
transfer vector {\bf Q} = (0.7, 0, 36.6) as a function of incident photon
energy $E_i$ above the corresponding edge. We found that the low energy fine
structure corresponding to intra-{\tg} excitations steadily decreases when the
incident photon energy changes from 11215 to 11220 eV.

\paragraph{\bf{O $K$ RIXS spectrum}}

Figure \ref{rixs_O_K_BNIO} shows the theoretically calculated RIXS spectrum at
the O $K$ edge in Ba$_4$NbIr$_3$O$_{12}$ in the GGA+SO approach together with
the partial contributions from different transitions between the energy
bands. The shape of the oxygen $K$ RIXS spectrum at 0$-$5 eV is very similar
to the corresponding RIXS spectrum at the Ir $L_3$ edge. The two peaks at
$\le$2 eV are derived from $O_{\tg} \rightarrow O_{\tg}$ transitions (the red
curve): the interband transitions between the oxygen 2$p$ states strongly
hybridized with the Ir {\tg} states in close vicinity to the Fermi level.  At
the same time, $O_{\tg} \rightarrow O_{\eg}$ transitions are very weak (the
orange curve).  We found that the major contribution to the fine structure
situated from 2 to 5 eV is derived from O$_{2p}$ $\rightarrow O_{\tg}$
transitions (the green curve), while O$_{2p}$ $\rightarrow O_{\eg}$
transitions are less intensive (the black curve).  Opposite to the Ir $L_3$
RIXS spectrum, the intensity of the oxygen $K$ RIXS spectrum for energies
$\ge$5 eV is very large. These structures are derived from 5$d_{\rm{O}}$
$\rightarrow O_{\tg}$ (the magenta curve), O$_{2p}$ $\rightarrow O_{\eg}$ (the
black curve), and O$_{\tg}$ $\rightarrow$ Ba$_{5d}$ transitions. Experimental
measurements of the RIXS spectrum at the O $K$ edge in Ba$_4$NbIr$_3$O$_{12}$
are highly desirable.

\paragraph{\bf{Nb RIXS spectra}}

\begin{figure}[tbp!]
\begin{center}
\includegraphics[width=1.\columnwidth]{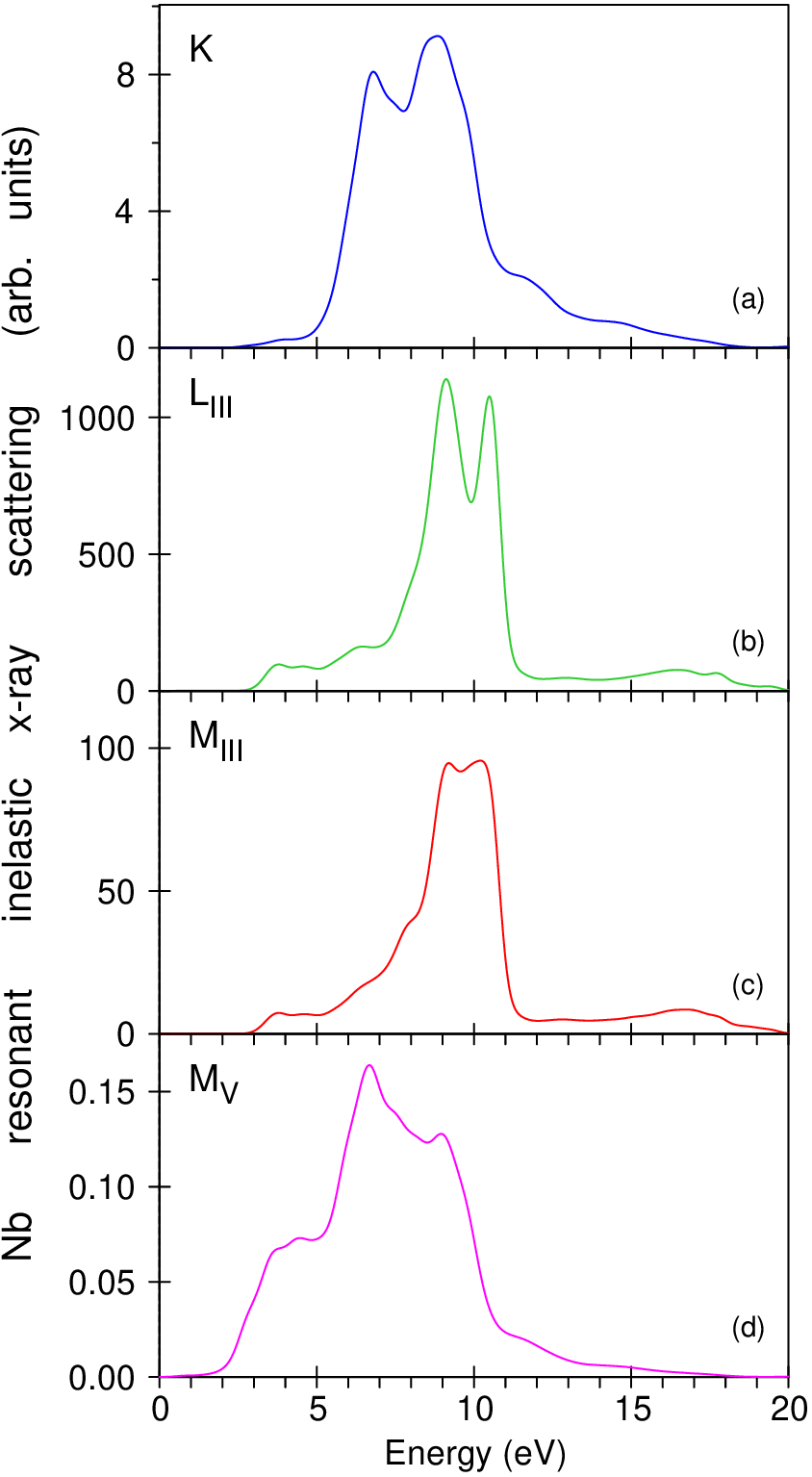}
\end{center}
\caption{\label{rixs_Nb_BNIO}(Color online) The theoretically calculated
  resonant inelastic x-ray scattering (RIXS) spectra at the Nb $K$ (a),
  $M_5$ (b), $L_3$ (c), $M_3$ (d), and $N_3$ (e) edges in
  Ba$_4$NbIr$_3$O$_{12}$. }
\end{figure}

At the core level RIXS is not only element-specific but also orbital
specific. For 3$d$ transition metals, the electronic states can be probed by
$K$, $L_{2,3}$ and $M_{2,3}$ x-ray absorption, emission, and RIXS spectra. For
4$d$ transition metals, the electronic states can be probed by $K$, $L_{2,3}$,
$M_{2,3}$, $M_{4,5}$, and $N_{2,3}$ spectra, whereas in 5$d$ transition metals
one can also use extra $N_{4,5}$, $N_{6,7}$, and $O_{2,3}$ spectra. For
unpolarized absorption spectra, transitions with $\Delta l= \pm 1, \Delta j=
0, \pm 1$ (one-particle dipole selection rules) are allowed only. Therefore,
only electronic states with appropriate symmetry contribute to the absorption,
emission, and RIXS spectra under consideration (Table \ref{tbl:dip_rule}). To
investigate the influence of the core state on the resulting Nb RIXS spectra
in Ba$_4$NbIr$_3$O$_{12}$ we calculated the RIXS spectra at the Nb $K$, $L_3$,
$M_3$, $M_5$, and $N_3$ edges. The results are presented in
Fig. \ref{rixs_Nb_BNIO}.

 \begin{table}[tbp]

 \caption{Angular momentum symmetry levels indicating the dipole allowed
   transitions from core states to the unoccupied valence states in transition
   metals with the partial density of states character.}

 \label{tbl:dip_rule}
 \begin{tabular}{cccccccccc}
\hline
\hline
&          &          &{$L_2$}   &{$L_3$}   &       &       &       &       &\\
&Spectra   &    K     &{$M_2$}   &{$M_3$}   &{$M_4$}&{$M_5$}&{$N_6$}&{$N_7$}&\\
&          &          &{$N_2$}   &{$N_3$}   &{$N_4$}&{$N_5$}&       &       &\\
&          &          &{$O_2$}   &{$O_3$}   &       &       &       &       &\\
 \tableline
&          &          &2$p_{1/2}$&2$p_{3/2}$&       &       &               &\\
&  Core    &1$s_{1/2}$&3$p_{1/2}$&3$p_{3/2}$&3$d_{3/2}$&3$d_{5/2}$&4$f_{5/2}$&4$f_{7/2}$&\\
&  level   &          &4$p_{1/2}$&4$p_{3/2}$&4$d_{3/2}$&4$d_{5/2}$  &       &\\
&          &          &5$p_{1/2}$&5$p_{3/2}$&       &       &               &\\
 \tableline
&Valence&$p_{1/2}$&$s_{1/2}$&$s_{1/2}$&$p_{1/2}$&$p_{3/2}$&$d_{3/2}$&$d_{5/2}$&\\
&states &$p_{3/2}$&$d_{3/2}$&$d_{3/2}$&$p_{3/2}$&$f_{5/2}$&$d_{5/2}$&$g_{7/2}$&\\
&       & -       & -     &$d_{5/2}$&$f_{5/2}$&$f_{7/2}$&$g_{7/2}$&$g_{9/2}$&\\
\hline
\hline
 \end{tabular}
 \end{table}

Figures \ref{rixs_Nb_BNIO}(a) and \ref{rixs_Nb_BNIO}(b) show the RIXS spectra
at the Nb $K$ and $M_5$ edges calculated in the GGA+SO approach.  The exchange
splitting of the initial $1s$ core state is extremely small
\cite{Ebe89,book:AHY04}, therefore only the exchange and spin-orbit splitting
of the 5$p$ states is responsible for the observed RIXS spectrum at the Nb $K$
edge. The Nb $K$ RIXS spectrum consists of two major peaks at 5.6 and 8.6 eV
energy. The low energy peak is due to the interband transitions from the
occupied Nb 5$p$ states hybridized with oxygen 2$p$ states to the empty Nb
5$p$ states hybridized with Ir {\tg} and {\eg} states.  The high energy peak
is due the transitions between Nb 5$p$ states and the states derived from the
hybridization with Ba 5$d$ states.

The Nb $M_5$ spectrum can be to some extent considered as an analog of the $K$
spectrum. The Nb $K$ RIXS spectrum reflects the energy distribution of
5$p_{1/2}$ and 5$p_{3/2}$ energy states (Table \ref{tbl:dip_rule}). The $M_5$
RIXS spectrum due to the dipole selection rules occurs for the transitions
from the 3$d_{3/2}$ core state to 5$p_{3/2}$, 4$f_{5/2}$, and 4$f_{7/2}$
states. Comparing this spectrum with the corresponding Nb $K$ RIXS spectrum
one can see an obvious resemblance between them.  The major difference is seen
from 0 to 7 eV. The difference between the Nb $K$ and $M_5$ spectra can be
attributed to an additional contribution of 4$f_{5/2,7/2}$ energy states to
the $M_5$ spectrum as well as to the difference in the radial matrix elements
($1s$ $\to$ 5$p_{1/2,3/2}$ for the $K$ spectrum and $3d_{5/2}$ $\to$
5$p_{3/2}$ for the $M_5$ spectrum). Although the latter plays a minor role due
to the fact that the radial matrix elements are smooth functions of energy
\cite{book:AHY04}.

The RIXS spectra in the row $L_3-M_3-N_3$ have almost a similar shape and
demonstrate a systematic decrease of intensity [Figures
  \ref{rixs_Nb_BNIO}(c,d,e)].  These spectra are more narrow than the $K$ and
$M_5$ RIXS spectra and shifted towards higher energy.  This can be explained
by a smaller width of occupied Nb 4$d$ states in comparison with the
corresponding Nb 5$p$ states. Besides, Nb 4$d$ states are situated below Nb
5$p$ states in energy (see Fig. \ref{PDOS_BNIO}).

\section{Conclusions}

To summarize, we have investigated the electronic structure of
Ba$_4$NbIr$_3$O$_{12}$ in the frame of the fully relativistic spin-polarized
Dirac approach. We have also presented comprehensive theoretical calculations
of the RIXS spectra at the Ir $L_3$, oxygen $K$, Nb $K$, $L_3$, M$_3$, $M_5$,
and $N_3$ edges as well as the XAS spectra at the Ir $L_{2,3}$ and Nb $L_3$
edges. To investigate the occupied valence states we have calculated the
photoemission spectrum in Ba$_4$NbIr$_3$O$_{12}$. The theoretically calculated
spectra are in good agreement with the experimental data.

Ba$_4$NbIr$_3$O$_{12}$ has a quasi-2D structure composed of corner-connected
Ir$_3$O$_{12}$ trimers containing three distorted face-sharing IrO$_6$
octahedra. The Ir atoms are distributed over two symmetrically inequivalent
sites: at the center of the trimer (Ir$_1$) and its two tips (Ir$_2$). The
Ir$_1$ $-$ Ir$_2$ distance within the trimer is quite small and equals to
2.547 \AA. As a result, there is clear formation of bonding and antibonding
states. The large bonding-antibonding splitting stabilizes the
{\dzz}-orbital-dominant antibonding state of 5$d$ holes and produces a wide
energy gap at the Fermi level. The Ir {\tg} bands are found to split into
distinct states with mixed character and these different states avoid crossing
one another, suggesting the formation of molecular orbital-like mixed orbitals
within the Ir$_3$O$_{12}$ trimers. Additional evidence of the QMO picture is
that the ground state of Ba$_4$NbIr$_3$O$_{12}$ is a nonmagnetic singlet. The
theoretically calculated BR in Ba$_4$NbIr$_3$O$_{12}$, which can be considered
as a direct probe of SOC, is equal to 2.72 for the Ir$_1$ site and 2.95 for
the Ir$_2$ site. It is smaller than for strongly spin-orbit coupled iridates,
such as Sr$_2$IrO$_4$. Relatively moderate SOC together with molecular
orbital-like states in Ba$_4$NbIr$_3$O$_{12}$ suggest that the pure
$J_{\rm{eff}}$ = 1/2 model may not be appropriate for this oxide.  It is
better use a description based on molecular orbital-like states.

The theoretically calculated Ir $L_3$ RIXS spectrum is in good agreement with
the experiment. We found that the low energy part of the RIXS spectrum $\le$2
eV corresponds to intra-{\tg} excitations. The peak located at $\sim$3.2 eV
was found to be due to $\tg \rightarrow \eg$ transitions. Some amount of $\tg
\rightarrow \tg$ transitions also contribute to the intensity of this
peak. The high energy fine structure $\ge$5.3 eV is mostly determined by
5$d_{\rm{O}}$ $\rightarrow \tg$ and O$_{2p}$ $\rightarrow$ {\eg}
transitions. The spectral features between 8 and 12 eV are due to
5$d_{\rm{O}}$ $\rightarrow \eg$ transitions. There is relatively large {\bf Q}
dependence of the RIXS spectrum at the Ir $L_3$ edge with pronounced periodic
modulation of the RIXS intensity as a function of Q$_z$. In our investigation
of the Ir $L_3$ RIXS spectrum as a function of incident photon energy $E_i$ we
found that the low energy fine structure corresponding to the intra-{\tg}
excitation steadily decreases when the incident photon energy changes above
the corresponding edge from 11215 to 11220 eV.

The shape of the oxygen $K$ RIXS spectrum at 0$-$5 eV is very similar to the
corresponding RIXS spectrum at the Ir $L_3$ edge. The two peaks at $\le$2 eV
are derived from the interband transitions between the oxygen 2$p$ states
strongly hybridized with the Ir {\tg} state in close vicinity to the Fermi
level.  We found that the major contribution to the fine structure situated
from 2 to 5 eV is derived from O$_{2p}$ $\rightarrow O_{\tg}$
transitions. Opposite to the Ir $L_3$ RIXS spectrum, the intensity of the
oxygen $K$ RIXS spectrum for energies $\ge$5 eV is very large. These
structures are derived from 5$d_{\rm{O}}$ $\rightarrow O_{\tg}$, O$_{2p}$
$\rightarrow O_{\eg}$, and O$_{\tg}$ $\rightarrow$ Ba$_{5d}$ transitions.

To investigate the influence of the core state on Nb RIXS spectra, we
calculated the RIXS spectra at the Nb $K$, $L_3$, $M_3$, $M_5$, and $N_3$
edges. We found a similarity between the Nb $K$ and $M_5$ RIXS spectra. The
major difference is seen from 0 to 7 eV.  The difference between the Nb $K$
and $M_5$ spectra can be attributed to an additional contribution of
4$f_{5/2,7/2}$ energy states to the $M_5$ spectra as well as to the difference
in the corresponding radial matrix elements. We found that the RIXS spectra in
the row $L_3-M_3-N_3$ have almost a similar shape and demonstrate a systematic
decrease of intensity.  These spectra are more narrow than the $K$ and $M_5$
RIXS spectra and shifted towards higher energy. This can be explained by a
smaller width of the occupied Nb 4$d$ states in comparison with the
corresponding Nb 5$p$ states. Besides, Nb 4$d$ states are situated below Nb
5$p$ states in energy. The RIXS spectroscopy of Nb atoms at the $K$, $L_3$,
$M_3$, $M_5$, and $N_3$ edges may be a very useful tool for the investigation
of the electronic structure of Ba$_4$NbIr$_3$O$_{12}$.

\section*{Acknowledgments}

The studies were supported by the National Academy of Sciences of
Ukraine within the budget program KPKBK 6541230 "Support for the development
of priority areas of scientific research".


\newcommand{\noopsort}[1]{} \newcommand{\printfirst}[2]{#1}
  \newcommand{\singleletter}[1]{#1} \newcommand{\switchargs}[2]{#2#1}

\end{document}